\documentclass[epj]{svjour}
\usepackage{graphics}
\usepackage[latin1]{inputenc}
\usepackage{epsfig}
\usepackage{a4wide}
\usepackage{latexsym}
\usepackage{amssymb}
\usepackage{amsmath}
\usepackage{graphics}
\usepackage{cite}
\def\d{{\rm d}}

\def\im{{\rm i}}

\def\x{{\bf x}}

\def\T{{\boldsymbol{\sf T}}}

\def\R{{\boldsymbol{\sf R}}}
\def\I{{\boldsymbol{\sf I}}}

\def\F{{\bf F}}
\def\A{{\bf A}}
\def\f{{\bf f}}
\def\k{{\bf k}}
\def\U{{\bf U}}
\def\V{{\bf V}}
\def\X{{\bf X}}
\def\Y{{\bf Y}}

\def\smalze{{\scriptscriptstyle{(0)}}}
\def\smalun{{\scriptscriptstyle{(1)}}}
\def\smaldu{{\scriptscriptstyle{(2)}}}
\def\smaln{{\scriptscriptstyle{(n)}}}
\def\smalCM{{\scriptscriptstyle{\rm CM}}}
\def\Pe{{\rm Pe}}
\def\beq{\begin{eqnarray}}
\def\eeq{\end{eqnarray}}
\begin{document}
\title{Effective swimming strategies
in low Reynolds number flows
}
\author{Piero Olla
}                     
%
%
\institute{ISAC-CNR and INFN, Sez. Cagliari, 
I--09042 Monserrato, Italy.}
\date{Received: date / Revised version: date}
%
\abstract{
The optimal strategy for a microscopic swimmer to migrate across
a linear shear flow is discussed. The two cases, in which the swimmer is
located
at large distance,
and in the proximity of a solid wall, are taken into account. It is shown that
migration can be achieved by means of a combination of sailing through
the flow and swimming, where the swimming strokes are
induced by the external flow without need of internal energy sources
or external drives. The structural dynamics required for the swimmer to
move in the desired direction is discussed and two simple models, based respectively
on the presence of an elastic structure, and on an orientation dependent friction,
to control
the deformations induced by the external flow, are analyzed. In all cases, the
deformation sequence is a generalization of the tank-treading motion regimes
observed in vesicles in shear flows.
Analytic expressions for the migration velocity as a function of the deformation pattern
and amplitude are provided. The effects of thermal fluctuations on propulsion
have been discussed and the possibility that noise be exploited 
to overcome the limitations
imposed on the microswimmer by the scallop theorem have been discussed.
} 
\authorrunning{Olla}
\titlerunning{Swimming strategies in low Reynolds number flows}
\maketitle
\section{Introduction}
\label{sec1}
There has been recently a resurgence of interest in low Reynolds number swimming.
Particular attention has been given to discrete designs 
\cite{purcell77,najafi04,avron05}, which allow simpler description of the geometrical
aspects of the problem, and identification 
of the necessary ingredients for propulsion.

One of the reasons for renewed  interest is
the progress in mechanical manipulation at the microscale, which has
allowed the realization of the first examples of artificial microscopic swimmers
\cite{dreyfus05,yu06,behkam06,leoni09}. These examples constitute
the first step towards the construction of ''microbots'', whose application
would be wide-spread, e.g. in medicine, as microscopic drug carriers
in not otherwise accessible regions of the human body. At the present
stage, however, most of such artificial swimmers are driven by
external fields, and the problem of an autonomous power source 
remains under study. 

Several solutions to this problem have been proposed.
Among them, various methods of rectification of Brownian motion
\cite{lobaskin08,golestanian09},
and mechanical reactions in the swimmer  
body, induced by inhomogeneity in the environment, e.g.
presence of a chemical gradient \cite{golestanian05,paxton06,pooley07}.

It has recently been suggested, that a microswimmer 
may extract the energy needed for locomotion, out of the 
velocity gradients in an external flow \cite{olla10}.
Based on a discrete design, that is a generalization of the
three-bead swimmer of
\cite{najafi04,earl07,golestanian08}, it was shown that
a microswimmer could migrate across a linear shear flow,
by a sequence of deformations induced by the external
flow itself. A continuous version of this microswimmer
has been described in \cite{olla10a}, based on the analogy
of the deformation sequence in the discrete case,
with the tank-treading motion
regime of a vesicle (or of a microcapsule) in a linear shear flow 
\cite{keller82,kraus96,barthes80}. 

Through tank-treading, a microscopic object such as a vesicle,
is able to maintain a fixed shape and orientation in an external
flow, with its surface circulating around its interior,
precisely as a tank-tread \cite{note}. 
The existence of a preferential shape 
and orientation for the object,
turns out to be one of the main ingredients for migration 
in an external flow. It should be mentioned that
tank-treading has already a long history, as a propulsion candidate
for swimming in quiescent fluids
\cite{purcell77,leshansky08}. Other examples of migration induced by
simultaneous rotation and deformation of solid objects in viscous
shear flows, have been analyzed both 
theoretically \cite{watari09} and experimentally \cite{marcos09}.
The design of a microswimmer that uses its internal 
degrees of freedom to spin rather than to swim, has been proposed
in \cite{dreyfus05a}.

The interesting aspect of external flow aided propulsion, 
is that the migration velocity scales linearly
in the stroke amplitude. 
This behavior is not surprising: the 
migration velocity of tank-treading vesicles in wall bounded flows, scales 
linearly with the deviation from spherical shape 
\cite{olla97,abkarian02,olla00,danker09}. 
In contrast, the velocity of a microswimmer
in a quiescent fluid, due to the constrains imposed by the scallop theorem 
\cite{purcell77}, is characterized by quadratic scaling
\cite{shapere89}. 

In the present paper, the analysis in \cite{olla10}, which focused
on the behavior of a discrete microswimmer in an infinite domain, will be extended
to the case of a wall bounded flow. Particular attention will be given
to identification of the deformation patterns associated with migration in 
different flow conditions, and with energy extraction from the flow.
A simple semi-quantitative analysis of the effect of thermal fluctuations 
will be provided as well, along the lines of the approach described
in \cite{dunkel09}.

The generation of specific deformation patterns,
requires the presence of a control 
system, modulating the response of the microswimmer to the
external flow (after all, this is what characterizes a microswimmer, as compares
to, say, a simple vesicle). The possibility of control through modification 
of the swimmer structural properties will therefore be examined, and
a simple example of control, by braking on the swimmer moving parts,
will be described in detail.

The paper is organized as follows. In Sec. \ref{sec2}, the basic 
equations of the model are presented and in Sec. \ref{sec2.1}, the
results in \cite{olla10} are briefly summarized.
In Sec. \ref{sec3}, the analysis is extended to the case of
a wall bounded flow.  In Sec. \ref{sec4},
the mechanism of energy extraction from the flow is discussed.
The structural dynamics of the swimmer is discussed in 
Sec. \ref{sec5}.  In Sec. \ref{sec5.1} an hypothesis of control system
to achieve migration is discussed.
In Sec. \ref{sec5.2}, the effect of thermal noise on passive swimming is
analyzed. Section \ref{sec6} is devoted to conclusions.
Technical details on the swimmer behavior in a wall bounded flow
are provided in the Appendix.

\section{Swimming strategy in free space}
\label{sec2}
\begin{figure}
\begin{center}
\includegraphics[draft=false,width=4.cm]{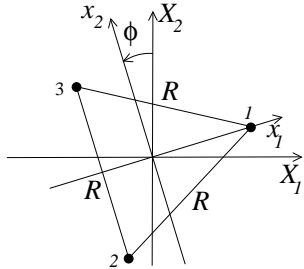}
\caption{
Rest configuration of the three-bead swimmer. 
Small case indicates the reference frame rotating
solidly with the device.
}
\label{orient}
\end{center}
\end{figure}

We analyze the behavior of the simple swimmer depicted in Fig. \ref{orient}.
Contrary to \cite{najafi04,golestanian08}, 
who considered a linear device, the three beads in the swimmer
under study are located, at rest, at the vertices of an equilateral triangle
of side $R$. 
The swimmer is placed in in a linear shear flow
\beq
\bar\U(\X)= \bar\U(0)+(0,\alpha X_1,0),
\label{eq1}
\eeq
and wants to migrate along the gradient direction $X_1$.

We are interested in a situation in which the system is able to move
by internal deformation without the aid of external forces.
We assume the links between the beads in the trimer to be immaterial
and the bead radii $a$ to be much smaller than $R$.
Including terms up to $O(a/R)$, the equation of motion
for the device can be written in the form:
\beq
\dot\X_i=\bar\U(\X_i)+\tilde\U_i(t)+\F_i(t)/\sigma,
\label{eq2}
\eeq
where $\X_i$ is the coordinate of the $i$-th bead, $\F_i(t)$
is the force on bead $i$ by the rest of the trimer,
$\sigma=6\pi \mu a$, with $\mu$ the fluid viscosity, is the
Stokes drag, and $\tilde\U_i(t)$ is
the flow perturbation in $\X_i$ generated by the other beads
in the trimer. 

To lowest order in $a/R$, the flow perturbation is obtained replacing the beads 
by point forces in the fluid, with intensity equal to the Stokes drag exerted
by the beads (stokeslet approximation \cite{happel}): 
\beq
\tilde\U_i(t)&=&\sum_{j\ne i}\T(\X_i-\X_j)\F_j;
\label{eq6}
\\
\T(\X)&
=&\frac{3a}{4\sigma}\Big[\frac{{\bf 1}}{|\X|}+\frac{\X\X}
{|\X|^3}\Big],
\label{eq7}
\eeq
where $\T(\X_i-\X_j)$, for $i\ne j$, is the off-diagonal part of the 
so called Oseen tensor \cite{happel}.

Linearity of the shear, Eq. (\ref{eq1}), and absence of external
forces, $\sum_i\F_i=0$, imply that the trimer center of mass
$\X^\smalCM=(\X_1+\X_2+\X_3)/3$ would move, if one disregarded
the flow perturbation, as a point tracer at $\X^\smalCM$:
$\dot\X^\smalCM=\bar\U(\X^\smalCM)+(1/3)\sum_i\U_i(t)$. Time averaging
the deviation with respect to $\bar\U(\X^\smalCM)$, we obtain the 
migration velocity
\beq
\U^{migr}=(1/3)\sum_i\langle\tilde\U_i\rangle=\langle\tilde\U_1\rangle.
\label{eq3}
\eeq
The velocity perturbation $\tilde\U_1$ is the sum of the stokeslet
fields $\tilde\U(\X|\X_i,\F_i)$, $i=2,3$ generated in $\X=\X_1$ by
beads $2$ and $3$. We assume exchange symmetry among the beads, 
in the sense that trimer configurations at angles $\phi+2n\pi/3$ are
indistinguishable. This mimicks the stationary tank-treading regime
of a vesicle, as the trimer arms will extend (or contract) in identical
way, when reaching a given position in the laboratory reference frame.
Exchange symmetry
allows us to rewrite Eq. (\ref{eq3}) in 
the equivalent form:
$\U^{migr}=\langle[\tilde\U(\X_2|\X_1,$ $\F_1)+\tilde\U(\X_3|\X_1,\F_1)]\rangle$.

\begin{figure}
\begin{center}
\includegraphics[draft=false,width=6.2cm]{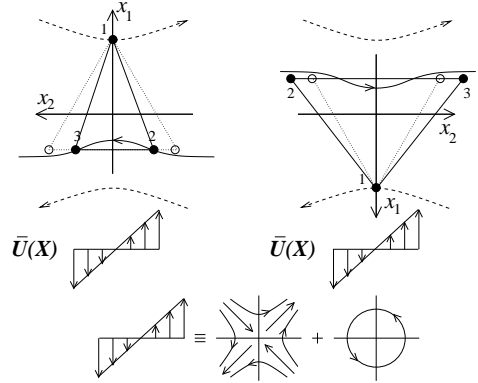}
\caption{The swimming strategy: link $23$ is contracted at $\phi=\pi/2$ (top-left)
and stretched at $\phi=-\pi/2$ (top-right); this causes beads 2 and 3 to sample
the stokeslet field by bead 1 [continuous line; see Eqs. (\ref{eq6}-\ref{eq7})] 
differently in the two configurations. The unbalance between the two configurations,
caused by deformation, is responsible for migration.
Dashed lines indicate the strain components
of $\bar\U$ producing the stokeslet in the two cases. Bottom figure:
decomposition of $\bar\U(\X)$ into strain and vorticity component.
}
\label{nuoto}
\end{center}
\end{figure}

Focusing on the effect of stokeslet 1 on beads 2 and 3, rather
than on the one of stokelets 2 and 3 on bead 1, allows to understand
geometrically the swimming strategy of the trimer,
as illustrated in Fig. \ref{nuoto} (we consider for the moment 
the case of an unbounded domain).

As shown in \cite{olla10} [see Eq. (\ref{eq13}) below], the undeformed trimer will rotate
in the field $\bar\U(\X)$ with constant frequency, equal to the flow vorticity $\alpha/2$.
Because of this, to lowest order in the deformation, $\tilde\U(\X|\X_1,\F_1)$ will
be the stokeslet field generated by action of the strain component of $\bar\U$ on a 
(fixed) bead at position $\X_1$. 
We see that the the $X_1$ component of the stokeslet field generated by bead 1 in 
response to the strain component
of $\bar\U$ is negative or positive depending on whether
$0<\phi<\pi$ or $\pi<\phi<2\pi$. 
Due to the approximately constant rotation frequency,
the trimer will spend an equal time at the two orientations indicated in figure,
and, in the case of the undeformed trimer, 
the contributions to $U_1^{migr}$ from the two orientations would cancel by symmetry.
Deformations, however, break this symmetry and
beads 2 and 3 will sample components $\tilde U_1(\X_{2,3}|\X_1,\F_1)$ that are in general
of different amplitude. The end result is
$\langle\tilde U_1(\X_{2,3}|\X_1,\F_1)\rangle\ne 0$, which 
leads to migration along $X_1$. 

Quantitative analysis [see Eq. (\ref{eq17}) below] shows
that the deformation sequence illustrated in Fig. \ref{nuoto} will lead to migration 
of the trimer to positive $X_1$ (to the right in figure).

We notice at this point that migration would be achieved also if just one arm 
in the trimer
performed the stretching-contraction sequence in Fig. \ref{nuoto}. [For small 
deformations, it is possible to see that the migration 
velocity would just be reduced by $1/3$ with respect to Eq. (\ref{eq3})].
We thus see that the presence of an external flow makes
one degree of freedom sufficient for locomotion. We recall that the scallop
theorem would prevent this, in the case of a microswimmer in a quiescent
fluid \cite{purcell77}. Similar ``violations'' of the scallop theorem were
observed in \cite{alexander08,lauga08}, in which case, the role of the external flow
was played by the perturbation generated in the fluid by other swimmers.
We shall see in Sec. \ref{sec5.2}, how a similar result could be obtained exploiting
the presence of thermal fluctuations.

\section{Migration in free space; quantitative theory}
\label{sec2.1}
To analyze the deformation dynamics of the trimer, it is 
convenient to work in the rotating reference;
Small case will identify vectors measured in the rotating reference frame
(see Fig. \ref{orient}). 
In the absence of rotational diffusion, the motion of the trimer will be the 
sum of a translation and a rotation in the plane $X_2X_3$, with rotation frequency
$\Omega=\dot\phi$, where $\phi$, as indicated in Fig. \ref{orient},
is the angle between the two reference frames.

\begin{figure}
\begin{center}
\includegraphics[draft=false,height=3.cm]{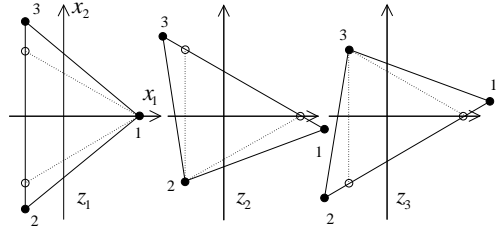}
\caption{
Deformations of the trimer corresponding to $z_i>0$ for $i=1,2,3$. In the three cases
$z_j=0$ for $j\ne i$.
}
\label{deform}
\end{center}
\end{figure}

We are interested in a regime of small deformations, and will proceed perturbatively
in the deformation amplitude. 
In particular, we will have, for the bead position in the rotating frame:
$\x_i=\x_i^\smalze+\tilde\x_i$,
with $\x_i^\smalze$ giving the bead positions for the undeformed trimer:
$\x^\smalze_1/R=(1/\sqrt{3},$ $0,0)$,
$\x_2^\smalze/R=-(1/(2\sqrt{3}),1/2,0)$, $\x_3^\smalze/R$ $
=(-1$ $/(2\sqrt{3}),1/2,0)$,
and $\tilde\x_i$ accounting for the deformations. We can express the trimer
deformations in terms of three arbitrary independent parameters $z_i$
(that with two translational and one rotational degrees of freedom make up
the six-dimensional configuration space of the trimer in the plane $X_1X_2$).
With the choice 
\begin{eqnarray}
\tilde\x_1&=&\frac{R}{2}\Big(\sqrt{3}(z_2+z_3),z^\smaln_3-z_2,0\Big),
\nonumber
\\
\tilde\x_2&=&\frac{R}{2}\Big(-\sqrt{3}z_3,-2z_1-z_3,0\Big),
\label{eq10}
\\
\tilde\x_3&=&\frac{R}{2}\Big(-\sqrt{3}z_2,2z_1+z_2,0\Big),
\nonumber
\end{eqnarray}
the deformation corresponding to just one $z_i$ being non-zero, will describe
stretching/contraction of the link opposite to bead $i$, while bead $i$ remains
fixed in the plane $x_1x_2$ (see Fig. \ref{deform}).
Assuming a stationary dynamics for the trimer, with 
exchange symmetry between the beads, we can write:
\beq
z_i=\sum_n[A_n\cos n\phi_i+B_n\sin n\phi_i],
\label{eq10.1}
\eeq
where $\phi_1=\phi$, $\phi_{2,3}=\phi\mp 2\pi/3$. 
It will be expedient in calculating the energy balance for locomotion to
consider a hierarchy of deformations of increasing order in some small parameter.
We thus write $z_i=z_i^\smalun+z_i^\smaldu+\ldots$,
with 
the normalized deformation amplitude
$z=\langle\sum_i(z_i^\smalun)^2\rangle$
playing the role of expansion parameter for the theory,
Superscripts will be used to indicate the order in $z$ at 
which a quantity is evaluated.

Perturbative analysis of Eq. (\ref{eq2}) using Eqs. (\ref{eq1}) and (\ref{eq10}) gives
the result \cite{olla10}:
\beq
\f_1^\smalze
=\frac{\alpha\sigma R}{\sqrt{3}}\Big(-\frac{1}{2}\sin 2\phi,-\frac{1}{2}\cos 2\phi\Big).
\label{eq14.1}
\eeq
and, to next order
\beq
f_{11}^\smalun&=&\frac{\alpha\sigma R}{2}\Big\{\frac{\sqrt{3}}{2}[{z^\smalun_2}'+{z^\smalun_3}'-
(z^\smalun_2+z^\smalun_3)
\nonumber
\\
&\times&\sin 2\phi]-\frac{1}{2}(z^\smalun_3-z^\smalun_2)\cos 2\phi\Big\},
\nonumber
\\
f_{12}^\smalun&=&
\frac{\alpha\sigma R}{2}\Big\{\frac{1}{2}
({z^\smalun_3}'-{z^\smalun_2}')
\nonumber
\\
&-&\frac{1}{\sqrt{3}}(z^\smalun_1+z^\smalun_2+z^\smalun_3)\cos 2\phi\Big\},
\label{eq15}
\eeq
where primes indicate derivative with respect to $\phi$.
Notice that we have disregarded the contribution from 
the velocity perturbation $\tilde\U_i$ to the force in Eq. (\ref{eq2}).
Clearly, $\f_1^\smalze$ is the reaction force of the rigid trimer against the external flow, 
and $\f^\smalun_1$ accounts for deformation.
Notice that the only contribution strictly qualifying as
swimming is the one proportional to $\dot\x_1^\smalun$ in $\f^\smalun$ [the ${z_{2,3}}'$ terms
in Eq. (\ref{eq15}); see also Eq. (\ref{eq10.1})], 
the remnant being better described as ``sailing''.

The same analysis leading to Eq. (\ref{eq15}) gives for the rotation frequency \cite{olla10}:
\beq
\Omega&=&\frac{\alpha}{4}\{2+ [\sqrt{3}\,(z^\smalun_2 -z^\smalun_3)\sin 2\phi
+(z^\smalun_2
\nonumber
\\
&+&z^\smalun_3-2z^\smalun_1)\cos 2\phi]+O(z^2)\},
\label{eq13}
\eeq
and from stationarity of the system, we can replace time averages with angular averages:
\beq
\langle h\rangle=\frac{1}{\pi\alpha}\int_0^{2\pi}\Omega(\phi)h(\phi)\d\phi.
\label{eq13.1}
\eeq
Expanding $\Omega$ in powers of $z$, we can then write the average in the above equation as
a sum of contributions of increasing order in $z$:
$\langle. \rangle=\langle .\rangle^\smalze+\langle .\rangle^\smalun+\ldots$,
with $\langle f\rangle^\smaln=(\pi\alpha)^{-1}\int_0^{2\pi}\Omega^\smaln(\phi)f(\phi)\d\phi$.

Knowledge of the force in the rotating reference frame, allows to write for the migration velocity,
substituting  Eq. (\ref{eq7}) into Eq. (\ref{eq3}):
\beq
\U^{migr}=\langle\R\T_1\f_1\rangle,
\label{eq8}
\eeq
where $\T_1=\T(\x_2-\x_1)+\T(\x_3-\x_1)$ and
$\R$ 
is the rotation matrix back to the laboratory frame:
\beq
R_{11}=R_{22}=\cos\phi;
\quad 
R_{21}=-R_{12}=\sin\phi.
\label{eq9}
\eeq
A simple calculation, including terms up to $O(z$ $\times a/R)$ gives for the Oseen tensor 
$\T_1$: $T_1^{11}=\beta\{7/2-[13z_1+29(z_2+z_3)]/8\}$, $T_1^{12}=T_1^{21}=
\beta(\sqrt{3}/8)(z_2-z_3)$ and
$\tilde T_1^{22}= \beta\{5/2+[2z_1-31(z_2+z_3)]/\}8$, with 
$\beta=3a/(4\sigma R)$. Substitution into Eq. (\ref{eq8}) and exploiting 
Eqs. (\ref{eq14.1}-\ref{eq13}),
gives the final result, to lowest order in $a/R$ and $z$:
\beq
U_1^{migr}=-\frac{\sqrt{3}\alpha a}{256}\Big[73B^\smalun_1+13B^\smalun_3\Big];
\label{eq17}
\eeq
thus $U^{migr}/(\alpha R)=O(za/R)$.
Let us consider the contribution $B_1\sin\phi_1$.
We see from Fig. \ref{nuoto},
that the stokeslet field generated by bead 1 in response to the strain component
of $\bar\U$,
has a positive or negative component at beads $2,3$, depending on whether
$0<\phi<\pi$ or $\pi<\phi<2\pi$. Migration is produced by the deformation induced
symmetry breaking between the two orientations.

\section{Migration in the presence of a wall}
\label{sec3}
A solid wall bounding the flow provides the swimmer with an additional mechanism 
for migration. The flow perturbation by the swimmer will be the superposition
of what would be produced in free space, and a wall correction that can be 
expressed as a sum of images and counterimages
of the freee-space perturbation, generated alternatively at the wall and at the surface
of the beads \cite{happel}.  In the stokeslet approximation of Eqs. (\ref{eq2}) and
(\ref{eq6}-\ref{eq7}), only the first image has to be taken into considerations.
 
\begin{figure}
\begin{center}
\includegraphics[draft=false,width=7.3cm]{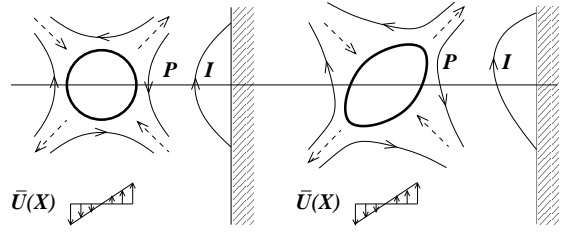}
\caption{The lift on an elongated object in a wall bounded flow. Dashed
arrows indicate the strain component of $\bar\U(\X)$.
In the case
of a sphere, the flow perturbation $P$ and of the image $I$ are reflection 
invariant around the plane perpendicular to the flow passing at the sphere center,
and the transverse drift is zero.
In the case of an elongated object, with the long axis along
the stretching direction of the strain, the lobes of the
flow perturbation and of the image will be tilted upwards, and the image field
at the object position will have a net component to the left. The opposite
will occur for an orientation at $\pi/2$ with respect to the one in figure.
}
\label{wall}
\end{center}
\end{figure}

The migration mechanism is illustrated in Fig. \ref{wall}, in the case
of an elongated object: the image at the wall of the free-space perturbation,
because of the asymmetry of the configuration, has a component
at the object center, that pushes it away from the wall. 
Now, a rigid object, with the exception of the very elongated structures described in 
\cite{bretherton62},
will rotate because of the flow vorticity. Because of this, such an object 
will typically alternate between a condition of outward and inward drift with respect to the
wall, and the transverse migration velocity will be zero.
 
In the case of deformable objects, e.g. vesicles, a fixed orientation and a non-zero transverse
drift can be achieved by means of tank-treading \cite{keller82}.
 
In the case of the triangular trimer, tank treading could be mimicked by
means of cyclic contraction of its arms: the arms contract when rotation leads them 
into the contracting (expanding) quadrant of the external strain. This will produce
an overall elongated shape oriented along the expanding (contracting) strain
direction, and lead to a non-zero transverse migration velocity.
To determine the contribution to
migration from presence of the wall, it is necessary to determine the image of
the stokeslet field of the beads in the trimer.
 
Let us suppose the wall to be located at coordinate $X_1=L$ with respect to the
trimer center of mass.
Let us denote by $\tilde\U^I(\X|\X_i,\F_i)$ the image of the stokeslet field
$\tilde\U(\X|\X_i,\F_i)$ 
generated in free space by the $i$-th bead.
In Eq. (\ref{eq8}), we thus have to add 
a wall contribution: 
$$
\Delta\U^{migr}=(1/3)\sum_{ij}\langle\tilde\U^I(\X_j|\X_i,\F_i)\rangle.
$$
For small $R/L$, we can Taylor expand $\tilde\U^I(\X_j|\X_i,$ $\F_i)$ 
around $\X_{i,j}=0$. 
From linearity of low Reyn- olds number hydrodynamics, we can write 
$\tilde\U^I(\X$ $|\X_i,\F_i)=\I(\X|\X_i)\F_i$, with $\I$ some tensor, and 
the lowest order contribution in $R/L$ will be $\I(0|0)\langle\sum_i\F_i\rangle=0$.
As regards the first order terms, from $\sum_j\X_j=0$, we have,
identifying the three bead contributions with the one from bead 1: 
$\langle\sum_j\X_j\cdot\nabla\tilde\U^I(\X|0,\F_1)\rangle_{\X=0}=0$. We thus remain with a
wall contribution to migration:
\beq
\Delta\U^{migr}&=&
3\langle(\X_1\cdot\nabla)\tilde\U^I(0|\X,\F_1)\rangle_{\X=0}
\nonumber
\\
&+&O((R/L)^3),
\label{eq18}
\eeq
and, from $\tilde\U^I(\X|\X_i,\F_i)=\I(\X|\X_i)\F_i$, we expect a result
in the form 
$\Delta U_\alpha^{migr}=
H_{\alpha\beta\gamma}\langle X_{1\beta}F_{1\gamma}\rangle$
(summation over repeated vector indices is assumed).
 
In order to determine the coefficients $H_{\alpha\beta\gamma}$, 
we must know the image field in Eq. (\ref{eq18}).
The image field of a stokeslet induced by a solid wall bounding the flow
was calculated in \cite{blake71}. Its derivation is outlined for reference
in the Appendix. We find for the image field derivatives entering Eq. (\ref{eq18}):
\beq
\partial_{X_1}\tilde U^I_1(0|\X,\F_1)|_{\X=0}&=&-\frac{9aF_{11}}{16\sigma L^2},
\nonumber
\\
\partial_{X_2}\tilde U^I_1(0|\X,\F_1)|_{\X=0}&=&\frac{9aF_{12}}{32\sigma L^2}.
\label{eq22}
\eeq
Substituting into Eq. (\ref{eq18}), leads to the result:
\beq
\Delta U^{migr}_1
=
\frac{27a}{16\sigma L^2}\langle M_{\alpha\beta} x_{1\alpha}f_{1\beta}\rangle,
\label{eq23}
\eeq
where $M_{\alpha\beta}=
(-R_{1\alpha}R_{1\beta}+\frac{1}{2}R_{2\alpha}R_{2\beta})$.
From Eq. (\ref{eq9}):
\beq
M_{\alpha\beta}=
\begin{pmatrix}
-\frac{1}{4}-\frac{3}{4}\cos 2\phi, 
&-\frac{3}{4}\sin 2\phi
\cr
-\frac{3}{4}\sin 2\phi,
&-\frac{1}{4}+\frac{3}{4}\cos 2\phi
\end{pmatrix}.
\label{eq24}
\eeq
As in the case of Eq. (\ref{eq17}), we can check that,
in the absence of deformation, $\Delta U^{migr}_1=0$, and that the lowest order
contribution with respect to $z$ to Eq. (\ref{eq23}) is 
$\langle M_{\alpha\beta}[x^\smalze_{1\alpha}f^\smalun_{1\beta}
+x^\smalun_{1\alpha}f^\smalze_{1\beta}]\rangle^\smalze$.
From the expressions of $M_{\alpha\beta}$,
and $\f^\smalze_1$ [see Eqs. (\ref{eq14.1}) and (\ref{eq24})], 
we see that only harmonics of order 
2 and 4 in $z$ contribute to $\Delta U^{migr}_1$. 
Direct calculation using
Eqs. (\ref{eq10}-\ref{eq10.1}) and (\ref{eq14.1}-\ref{eq17}), gives in fact,
to lowest order in $z$ and $R/L$:
\beq
\Delta U^{migr}_1=(243B^\smalun_2-729B^\smalun_4)\frac{\alpha aR^2}{2048L^2}.
\label{eq25}
\eeq
Notice that the 
$O((R/L)^2)$ behavior in Eq. (\ref{eq25}) is that of the image of a stresslet field
at the trimer position \cite{happel} (the quadrupole field depicted in Fig. \ref{wall}).
As can be seen in Eq. (\ref{eq25}), it is possible that cancellations arise between
the $n=2$ and $n=4$ harmonics, in which case the next order in the expansion in $R/L$
for $\Delta U^{migr}$ should be taken into account. Now, the image field
of a stokeslet at the wall, can be expressed as a multipole expansion, whose lowest order terms
are associated with stokeslets placed at the other side of the wall. It is worth
mentioning, therefore, that cancellations similar to the one discussed above have 
occur also in the interaction of two linear swimmers \cite{pooley07a}.

The situation in Eq. (\ref{eq25}) 
corresponds to the picture of fixed orientation and migration induced by tank-treading
described at the beginning of this section.
To fix the ideas, consider $B_4=0$, and focus on the deformation associated with
$z_1$. The regime $B_2<0$ corresponds to migration away from the wall. At the same time,
for $B_2<0$, 
$\phi=\pm\pi/4$ will correspond
respectively to contraction and stretching of the side $23$ of the trimer. 
In other words, migration away from the wall corresponds to the trimer 
maintaining a deformed shape, with long axis
along the stretching direction of $\bar\U$. This is the same
behavior of a tank-treading vesicle in a wall bounded flow, in the limit
of zero viscosity of the internal fluid and zero membrane friction 
\cite{olla97,abkarian02}.

\section{Energy balance}
\label{sec4}
Let us calculate the energy needed to perform the swimming actions described
in Secs. \ref{sec2} and \ref{sec3}.
The average power
exerted by the swimmer on the fluid is $P=3\langle\dot\x_1\cdot\f_1\rangle$.
This provides a lower bound on the power actually expended by the swimmer,
which must include the contribution by internal friction.

Let us consider first the case of an ideal trimer.
The lowest order contribution to the expended power, from $\dot\x^\smalze_i=0$,
is $P^\smalun=
\langle\dot\x^\smalun\cdot\f^\smalze\rangle^\smalze$. Now, in the free-space case
of Eq. (\ref{eq17}): $\x_1^\smalun=\x_1^\smalun(z_i^{free})\equiv\x^{free}$,
where $z_i^{free}=B_1\sin\phi_i+B_3\sin 3\phi$ and $\x_1^\smalun(z_i^{free})$
is given by the first of Eq. (\ref{eq10}). Thus, while $\x^{free}$ is odd with respect to
$\phi$, the corresponding force
$\f^\smalze$ is even [see Eq. (\ref{eq14.1})] and $P^\smalun=0$
automatically. 

A simple calculation shows that $P^\smalun=0$ also 
for the wall
contribution of Eq. (\ref{eq25}). Also in this case, the mechanism is easy to 
understand: focusing
on the sequence of contraction and stretching of side 23, produced by
a deformation $z_1^{wall}=B_2\sin(2\phi)+B_4\sin(4\phi)$, 
we see that the elongation of side $23$
for $\phi$ going from $-3\pi/4$ to $-\pi/4$ (energy gained from the fluid)
is compensated by contraction in passing from $-\pi/4$ to $\pi/4$ (energy lost to the
fluid). In similar way it can be shown that the contribution to
$\langle \dot\x^{free}\cdot\f^\smalze\rangle$ and $\langle \dot\x^{wall}\cdot\f^\smalze\rangle$,
from $\Omega^\smalun$, in the angular average of Eq. (\ref{eq17}), is zero.

The power expended by an ideal swimmer in free space, would be 
therefore
\beq
P^{free}=3\langle\dot\x^{free}\cdot\f^{free}\rangle^\smalze+O(z^3),
\label{eq26}
\eeq
where $\f^{free}=\f^\smalun_1(z^{free}_i)$ [see Eq. (\ref{eq15})], and a similar 
equation is obeyed by the power $P^{wall}$ that would be expended by the
trimer in the case of a wall bounded flow (notice that 
$\langle\x^{free}\cdot
\f^{wall}\rangle^\smalze=\langle\x^{wall}\cdot\f^{free}\rangle^\smalze=0$).

In order for the constrain force $\f_1^\smalze$ to produce work, it is necessary
that $\x_1^\smalun$ has a component $\x^{extr}=\x_1(z_i^{extr})$, with $z_i^{extr}=
A_2\cos 2\phi_i$. 
From Eqs. (\ref{eq10}), (\ref{eq14.1}) and (\ref{eq15}), this would correspond to 
the trimer extracting from the fluid a power
\beq
P^{extr}=
-3\langle\dot\x^{extr}\cdot\f_1^\smalze\rangle^\smalze
=\frac{\alpha^2\sigma R^2}{2}A_2.
\label{eq27}
\eeq
To understand the mechanism of power extraction, let us
focus on the contraction and stretching of arm 23 produced by the
deformation $z_1^{extr}$. We see that a positive $A_2$, from Eq. (\ref{eq10.1}), corresponds to 
link 23 being stretched when it is parallel to the flow direction, and contracted when
it is perpendicular to it.
Energy extraction from the flow comes from the fact that the $|\X_{2,3}|$
increase or decrease depending on whether the respective beads lie in the stretching or contracting
quadrants of the external strain (see Fig. \ref{nuoto}). 

In principle, a swimmer could use the mechanism outlined above to extract energy from 
the flow and store it for later use, say, through a system of springs. In alternative,
this energy could be utilized to compensate the power dissipated in swimming, as
accounted for in Eq. (\ref{eq26}).

In the case of the
ideal trimer described in Eq. (\ref{eq26}): 
$P^{extr}=P^{free}+P^{wall}$, which gives
$z^{extr}=O(z^2)$,
and $A_2=A_2^\smaldu+O(z^3)$
Internal friction, however, may contribute to dissipated power to $O(z)$,
and a deformation component $z^{extr}$ of the same amplitude as the swimming stroke
[see Eqs. (\ref{eq17}) and (\ref{eq25})] would in this case be required.
As it will be illustrated in the next section, this is going to be a rather 
natural situation, if some kind of elastic structure for the trimer is 
assumed.
\section{Dynamics of the elastic trimer}
\label{sec5}
We would like to understand the structural dynamics of a trimer undergoing the 
deformations described in the previous sections.

Let us analyze first the behavior of an elastic trimer, in the absence of any internal
system of control of the device response to the flow.
Indicating with $\psi_i$ the angle between the arms joining at bead $i$ and with
$x_{ij}$ the length of arm $ij$,
the potential energy due to bending and to stretching can be written in the form
\beq
\Delta U_B&=&\frac{\kappa_B R^2}{2}\sum_i\Delta\psi_i^2,
\nonumber
\\
\Delta U_S&=&\frac{\kappa_S}{2}\sum_{i>j}\Delta x_{ij}^2,
\label{eq61}
\eeq
where $\psi_i=\pi/3+\Delta\psi_i$ and $x_{ij}=R+\Delta x_{ij}$; 
$\kappa_BR^2$ is bending elasticity of the joints between the
trimer arms and $\kappa_S$ is the stretching elasticity of the arms.
Stretching and bending as a function of $z_i$ are obtained
from Eq.  (\ref{eq10}):
\beq
&&\Delta x_{32}= (\frac{5}{2}z_1+z_2+z_3)R,
\nonumber
\\
&&\Delta\psi_1= \sqrt{3}[z_1-\frac{1}{2}(z_2+z_3)],
\label{eq62}
\eeq
and cyclic permutations.
Substituting into Eq. (\ref{eq61}), 
we find the expression for
bending and stretching energy:
\beq
\Delta U_B&=&\frac{9\kappa_B R^2}{4}[z_1^2+z_2^2+z_3^2
\nonumber
\\
&-&(z_1z_2+z_2z_3+z_3z_1)],
\nonumber
\\
\Delta U_S&=&\frac{\kappa_S R^2}{2}\Big[\frac{33}{4}(z_1^2+z_2^2+z_3^2)
\nonumber
\\
&+&12(z_1z_2+z_2z_3+z_3z_1)\Big].
\label{eq63}
\eeq
Energy balance requires that $\Delta U_B+\Delta U_S+\Delta W+\Delta W_{in}=0$
where $\Delta W=\sum_i\f_i\cdot\Delta\x_i$ is the work exerted by the trimer on the 
fluid and $\Delta W_{in}$ is the work against internal friction forces. As discussed
in the previous section, 
$\Delta W^\smalun$ $=\sum_i\f_i^\smalze\cdot\Delta\x_i^\smalun$ 
averages to zero in a cycle.
From Eqs. (\ref{eq10}),
(\ref{eq14.1}-\ref{eq15}) and (\ref{eq13.1}), we obtain:
\beq
\Delta W^\smalun=\frac{\alpha\sigma R^2}{2}\sum_iz_i\sin 2\phi_i.
\label{eq65}
\eeq
Differentiating $\Delta W^\smalun+\Delta U_B+\Delta U_S=0$ with respect to $z_i$, $i=1,2,3$, 
gives the force balance equation in the absence of dissipation:
\beq
&&(6\kappa_B+11\kappa_S)z_1+(8\kappa_S-3\kappa_B)(z_2+z_3)
\nonumber
\\
&&=-\frac{2\alpha\sigma}{3}\sin 2\phi_1
\label{eq66}
\eeq
and cyclic permutations for $\phi_{2,3}$. Let us assume for simplicity 
that the friction forces acting in the
trimer are linear in $\dot\psi_i$ and $\dot x_{ij}$. 
Including friction leads to an equation in the form
\beq
&&\frac{\alpha}{2}[(6\gamma_B+11\gamma_S)z'_1+(8\gamma_S-3\gamma_B)(z'_2+z'_3)]
\nonumber
\\
&&+(6\kappa_B+11\kappa_S)z_1
+(8\kappa_S-3\kappa_B)(z_2+z_3)
\nonumber
\\
&&=-\frac{2\alpha\sigma}{3}\sin 2\phi_1,
\label{eq68}
\eeq
where 
$\gamma_BR^2$ and $\gamma_S$ are bending and stretching friction coefficients. 
Notice that, if
$\gamma_{B,S}\sim\alpha\kappa_{B,S}$, internal friction will produce an $O(z)$ 
contribution to the dynamics and
energy dissipation in the fluid can be disregarded.

To solve Eq. (\ref{eq68}), we assume a solution in the form
$z_i=A_2\cos 2\phi_i+B_2\sin 2\phi_i$ and obtain, after little algebra: 
$
(\kappa A_2+\alpha\gamma B_2/2)\cos 2\phi_i+[\kappa B_2-\alpha\gamma A_2/2
+2\alpha\sigma/9]\sin 2\phi_i=0,
$
where $\kappa=3\kappa_B+\kappa_S$ and $\gamma=3\gamma_B+\gamma_S$. 
Setting the coefficients in front of $\cos 2\phi_i$
and $\sin 2\phi_i$ independently equal to zero gives $A_2=-\alpha\gamma/(2\kappa)\,B_2$ and
$B_2=-2\kappa\sigma\alpha/(9(\kappa^2+\alpha^2\gamma^2/4))$; in other words $B_2<0$ and
$A_2>0$. Notice that $B_2<0$ corresponds to the tank-treading regime with 
long trimer axis along the stretching direction of $\bar\U$ described in Sec. \ref{sec3},
while $A_2>0$ corresponds to the energy transfer from $\bar\U$ to the trimer dynamics
discussed in correspondence of Eq. (\ref{eq27}).

The solution to Eq. (\ref{eq68}) can be written in alternative form as
$z_i=-\sqrt{A_2^2+B_2^2}\sin(2\phi_i+\arctan A_2/B_2)$, i.e.:
\beq
z_i
&=&\frac{-2\alpha\sigma}{9\sqrt{\kappa^2+\alpha^2\gamma^2/4}}
\nonumber
\\
&\times&\sin\Big(2\Big(\phi_i-\frac{1}{2}\arctan\frac{\alpha\gamma}{2\kappa}\Big)\Big).
\label{eq69}
\eeq
With the aid of Figs. \ref{orient} and \ref{deform} we can understand the deformation 
pattern described by Eq. (\ref{eq69}), and 
notice the analogy with the behavior of a tank-treading vesicle 
\cite{keller82} or a microcapsule \cite{barthes80}
in a viscous shear flow.
In the absence of dissipation, the trimer would maintain its long axis aligned with
the stretching direction of $\bar\U$. Adding dissipation would cause 
the long axis to rotate towards the flow, and to get aligned with it in the limit
$\alpha\gamma/\kappa\to\infty$.
No transition to a 
tumbling regime exists. In order for such a transition to occur, it would 
be necessary that the trimer rest shape were not that of an equilateral triangle.
Notice that this is the behavior of a microcapsule whose rest shape is that of a sphere
\cite{barthes80}.

From the analysis in Sec. \ref{sec3}, we see that,
in the absence of an internal control system, 
the only migration pattern of our trimer, 
could be migration away from
a wall bounding the flow.

\section{Swimming through braking}
\label{sec5.1}
We have seen that the tank-treading regime of Eq. (\ref{eq69}), which leads to
migration away 
from a wall, is a condition that does not require the presence of a particular
control system in the trimer. Things change if we wish to implement the behaviors
leading to Eqs. (\ref{eq17}) and (\ref{eq25}), i.e. migration in an unbounded
flow and migration towards a wall.
It has been shown in \cite{olla10} that a simple
orientation dependent ''braking'' system is sufficient to produce the
deformation sequences required for migration. We want to analyze here
the energetics of the system.

For the sake of simplicity,
consider $\kappa_B=\kappa_S=0$, so that the dynamics is dominated by
friction, and set $\gamma_S=3\gamma_B/8$, so that the system of equations (\ref{eq69})
becomes diagonal. This is likely not to lead to maximum efficiency in terms 
of speed vs. deformation amplitude, but provides an example that is easier perhaps 
to implement experimentally, than variable strength springs at the
trimer links and joints.

Under the present hypotheses, Eq. (\ref{eq68}) becomes
\beq
\gamma(\phi_i)z'_i=-\frac{4\sigma}{9}\sin 2\phi_i.
\label{eq70}
\eeq
The two situations leading to drift away from a wall, and migration in an unbounded
flow would require
$z_i=B_2\sin 2\phi_i+\ldots$ with $B_2>0$ and $z_i=B_1\sin\phi_i+\ldots$, respectively.

The first situation could be realized with 
$\gamma=\gamma_0\,(1+c\sin 4\phi_i)^{-1}$, $0<c<1$. The ''brake''
is released while vertex $i$ has passed the direction of maximum stretching, 
it is acted on entering the contracting quadrant, and is released again after 
passing the direction of maximum contraction.  Substituting
into Eq. (\ref{eq70}) we get in fact:
\beq
z_i&=&\frac{2\sigma}{9\gamma_0}[\cos 2\phi_i
\nonumber
\\
&+&c\,(\frac{1}{4}\sin 2\phi_i+\frac{1}{12}\sin 6\phi_i)].
\label{eq71}
\eeq
The second situation could be realized instead with
$\gamma=\gamma_0\,(1+c\sin\phi_i)^{-1}$, $|c|<1$. In this case,
the brake acts when the vertex is in the direction of the flow and is released when it is oriented
opposite to it (or vice versa, if $c$ has opposite sign).
Substituting
into Eq. (\ref{eq70}), we get in this case:
\beq
z_i=\frac{2\sigma}{9\gamma_0}[\cos 2\phi_i+c\,(\sin \phi_i-\frac{1}{6}\sin 3\phi_i)].
\label{eq72}
\eeq
Notice in both Eqs. (\ref{eq71}) and (\ref{eq72}), 
the term $\propto\cos 2\phi_i$, that signals energy transfer from the
fluid to the trimer.
\section{The effect of thermal fluctuations}
\label{sec5.2}
Going to sufficiently small scales, thermal fluctuations will 
play an increasingly important role in the microswimmer dynamics
(see e.g.
\cite{lobaskin08,golestanian09,dunkel09}).

In the present passive swimming scheme, thermal noise can act in essentially
two ways: 
\begin{itemize}
\item Modification of the microswimmer orientation with respect to the external flow.
\item Direct contribution to the microswimmer deformation and hence to propulsion. 
\end{itemize}
The magnitude of the first effect depends on the ratio of the rotational diffusion 
time $\tau_f$ (called 
in \cite{dunkel09} flipping time) and the hydrodynamic timescale $\alpha^{-1}$.
(For the sake of simplicity we shall restrict the analysis to the case of a 
swimmer constrained to the plane $X_1X_2$).
The flipping time of the trimer is obtained from the translational
diffusivity $D$ of an individual bead suspended in the fluid, from the dimensional relation
$\tau_f\sim R^2/D$. The diffusivity of a bead of radius $a$ and density $\rho$,
in a fluid at temperature $T$,
can be written in the form
\beq
D
\sim\frac{KT}{\mu a}
\label{diffusivity}
\eeq 
where $K$ is the Boltzmann constant.
From here, we obtain for the flipping
time
\beq
\tau_f\sim\frac{R^2}{D}\sim\frac{\mu a R^2}{kT}
\eeq
and the condition of negligible rotational diffusion becomes
\beq
\alpha\tau_f\sim\frac{1}{\Pe}\Big(\frac{R}{a}\Big)^2\gg 1;
\quad
\Pe\sim\frac{D}{\alpha a^2}.
\label{peclet}
\eeq
The dimensionless number $\Pe$, that is the Peclet number for the bead,
parameterizes the relative importance of thermal diffusion and hydrodynamic
transport at scale $a$.

Let us pass to analyze the contribution of thermal fluctuations to the swimming
strokes. Let us decompose the deformation in deterministic and fluctuating components:
\beq
\delta R=\delta\bar R+\delta r.
\nonumber
\eeq
The condition of negligible noise contribution to the swimming strokes will be
therefore $\langle\delta r^2\rangle$ $\ll\langle\delta\bar R^2\rangle$.

The evolution of the deterministic component $\delta\bar R$ is governed by
Eqs. (\ref{eq66}) and (\ref{eq68}). Let us consider a situation in which elastic and
friction force in the trimer contribute at the same level to propulsion.
This implies a force balance:
\beq
\Gamma\delta\dot{\bar R}\sim\omega^2\delta\bar R\sim\frac{\alpha}{\tau_a}R\gg\delta\ddot{\bar R}
\label{balance}
\eeq
where $\Gamma\sim\gamma/m$, $\omega^2\sim\kappa/m$ and 
$\tau_a\sim m/\sigma$, with $m\sim\rho a^3$ the bead mass; $\kappa$ and
$\gamma$ are the (time dependent) elastic constant and friction coefficient of the
trimer arms; $\tau_a$ is the Stokes time of the beads. Notice that the force
balance Eq. (\ref{balance}) implies
\beq
\frac{\omega}{\alpha}\sim\frac{1}{\sqrt{S_az}}\gg 1;
\quad
\Gamma\tau_a\sim\frac{1}{z}\gg 1
\label{ordering}
\eeq
where $S_a=\alpha\tau_a\ll 1$ defines the bead Stokes number.

The evolution of the fluctuating component can be described in terms of a stochastic differential 
equation in the form \cite{dunkel09}:
\beq
\delta\dot r+\frac{\omega^2}{\Gamma}\delta r\sim D_a^{1/2}\xi,
\quad
D_a\sim\frac{D}{\Gamma\tau_a},
\label{fluctuating}
\eeq
where $\xi(t)$ is a normalized white noise term: $\langle\xi(t)\xi(t')\rangle=\delta(t-t')$. The
diffusion coefficient $D_a$ is determined from the equipartition condition
\beq
\kappa\langle\delta r^2\rangle\sim m\langle\delta \dot r^2\rangle\sim KT,
\nonumber
\eeq
and we can write $D/\tau_a\sim D_a\Gamma\sim v_{th}^2$,
where $v_{th}$ is the thermal speed of the bead in a fluid at temperature $T$.

From Eqs. (\ref{ordering}) and (\ref{fluctuating}), 
we see that the relaxation time for $\delta r$ is the hydrodynamic
time $\alpha^{-1}$, and the fluctuation amplitude is 
$\langle\delta r^2\rangle\sim D_a\Gamma/\omega^2$. From Eqs. (\ref{ordering}) and 
fluctuating, we find the condition of negligible fluctuating component in the 
swimming strokes:
\beq
\frac{\langle\delta r^2\rangle}{\langle\delta\bar R^2\rangle}
\sim\frac{1}{z}\Big(\frac{a}{R}\Big)^2\Pe\ll 1
\label{stroke}
\eeq
A similar approach could be adopted also in the absence of an elastic component in the
trimer dynamics. In this case, the condition of negligible deformation fluctuations
would take the form $D_a\alpha^{-1}\ll\langle\delta\bar R^2\rangle$, corresponding
to the requirement that diffusion in $\delta r$ be small at hydrodynamic
timescales. It is possible to see that this leads again to the result in 
Eq. (\ref{stroke}).

Equations (\ref{peclet}) and (\ref{stroke}) provide the necessary conditions for the 
noise-free analysis in the previous section to be valid. 
We have to verify that
the resulting migration [Eq. (\ref{eq17})] is not overcome by diffusion. 

Over sufficiently small distances, the trimer migration will always have
a diffusive character. The degree of noisiness of the trimer migration 
could be parameterized in term of the ratio of the crossover distance
$R_c$ above which the trimer trajectory becomes ballistic, and the trimer
size $R$. This crossover is easily shown to occur at $R_c\sim D/U^{migr}$, corresponding
to a crossover time $\tau_c\sim R_c/U^{migr}$. (We are disregarding
the contribution to diffusivity from random swimming, which can be shown
to be appropriate provided the geometric constraints $z,a/R\ll 1$ are
satisfied \cite{note1}).
From Eq. (\ref{eq17}),
we have $U^{migr}\sim\alpha a\delta R/R$, and, using Eqs. (\ref{diffusivity}-\ref{stroke}),
we obtain
\beq
\frac{R_c}{R}\sim\Big(\frac{\tau_c}{\tau_f}\Big)^2\sim\Pe\frac{a}{zR}\ll \Pe^{1/2},
\label{crossover}
\eeq
and $R_c\ll R$ provided $\Pe$ is not too large and the conditions in Eqs. (\ref{peclet}) 
and (\ref{stroke}) are satisfied.
This crossover in space corresponds to a crossover in time, which, as described by
the relation
\beq
\alpha\tau_c\sim\Pe/z^2
\label{alphatau_c}
\eeq
[see Eqs. (\ref{peclet}) and (\ref{crossover})],
may occur at $\tau_c\gg\alpha^{-1}$ or $\tau_c\ll\alpha^{-1}$,
depending on the relative size of the parameters $\Pe$ and $z$.

Comparing Eqs. (\ref{peclet}-\ref{crossover}), we see that the effect of noise
on propulsion 
can be disregarded, if either (or both) $\Pe$ and $a/\delta R$ are very small.

\subsection{Large thermal fluctuations and the scallop theorem}
From Eqs. (\ref{diffusivity}) and (\ref{peclet}), we see that the Peclet number
$\Pe$ becomes $O(1)$ for values of $a$ in the range of the micrometer, and 
shear strengths of the order of $1/{\rm s}$. The transition to a diffusivity
dominated regime is rather sharp due to the inverse cubic dependence of $\Pe$ on the
bead size $a$. A trimer with submicron beads, would therefore diffuse in the
fluid without being able to exploit the presence of an external flow; in order
to achieve propulsion, if migration by swimming remains the strategy,
some internal motor would be required to overcome bead diffusion.

It is interesting to see that thermal noise could be exploited
to allow the swimmer to propel itself with just one degree of freedom available.
The mechanism
turns out to be not particularly efficient.  
Nevertheless, it provides another instance of how the constraints of 
the scallop theorem could be bypassed if some additional means (in this case rotational
diffusion) allows the swimmer to change orientation between one stroke and its reverse.

\begin{figure}
\begin{center}
\includegraphics[draft=false,width=7.cm]{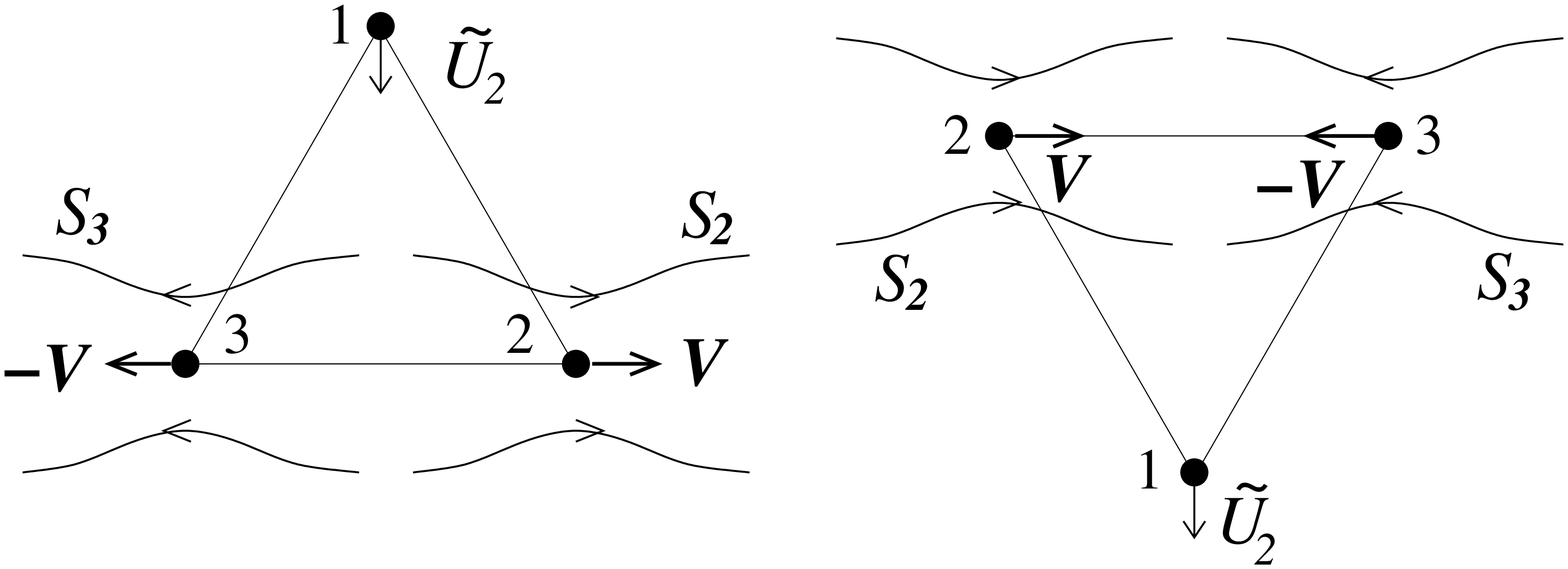}
\caption{
Active swimming in a noisy environment. Thermal fluctuations cause the trimer to
change continuously orientation. When bead 1 is pointing upward (left) an internal motor
stretches link 23; the opposite occurs when bead 1 is pointing downward (right);
$S_{2,3}$ indicate lobes of the stokeslet generated by beads 2 and 3.
The end result is a velocity perturbation $\tilde U_1$ at bead 1, with a
net component pointing downwards independent of orientation.
}
\label{noise}
\end{center}
\end{figure}

The mechanism is described in Fig. \ref{noise}.
An internal motor stretches and contracts one link in the trimer, when 
the opposite bead is backward or forward oriented with respect to the 
desired direction of motion.

To fix the ideas, suppose that the fixed frame is oriented with the $X_2$ axis pointing
upward in figure, so that $\tilde U_1\equiv(0,\tilde U_{12},0)$. The migration 
velocity can be estimated from the velocity perturbation at bead 1 when link 23 is
parallel to the $X_1$ axis. If $\pm\V$ is the instantaneous velocity of beads 2 or 3
in the deformation, the perturbation at $\X_1$ will be the sum of the two 
stokeslet fields
$\tilde\U(\X_1|\X_{2,3},\F_{2,3})=\tilde\U(\X_1|\X_{2,3},\pm\sigma\V)$.
The perturbation strength is obtained from 
Eqs. (\ref{eq2}-\ref{eq7}):
\beq
\tilde U_{12}=2T_{21}(\X_{13})F_{31}=-\frac{3\sqrt{3}}{8}\frac{a}{R}V,
\label{tildeUnoise}
\eeq
and the sign remains the same inverting the orientation. (The first index in $\tilde U_{12}$ 
and $F_{31}$ labels the bead; the two indices in $T_{21}$ are vectorial).

As anticipated, this swimming strategy is not very efficient.
The swimming strokes must be carried out with a characteristic frequency $\tau_f^{-1}$
to exploit rotational diffusion. Hence:
\beq
V\sim\frac{\delta R}{\tau_f}\sim\frac{D\delta R}{R^2}
\nonumber
\eeq
which gives, from Eq. (\ref{tildeUnoise}):
\beq
U^{migr}\sim\frac{Da\delta R}{R^3}
\nonumber
\eeq
Combining with Eq. (\ref{crossover}), we obtain however:
\beq
\frac{R_c}{R}\sim\frac{D}{U^{migr}R}\sim\frac{1}{z}\frac{R}{a}.
\label{inefficient}
\eeq
Unless $R\sim a$ and $\delta R/R$ is not too small (this would require a continuous swimmer 
undergoing large amplitude deformations) our device will have to swim a lot, before 
migration is able to overcome diffusion.
\section{Conclusion}
\label{sec6}
We have analyzed the behavior of a device that can swim by
extracting energy from the gradients in an external shear flow.
Adoption of a simple model, such as the three-sphere swimmer of 
\cite{najafi04}, has allowed to identify optimal swimming strategies,
both in infinite and wall bounded domains.

In order to migrate across a shear flow in an infinite domain, the microswimmer must
maintain on the average a configuration that is fore-aft asymmetric 
along the flow.
In order to migrate away from (towards) a wall perpendicular to the direction 
of the shear gradient, the microswimmer must maintain on the average
an elongated configuration along the stretching (contracting) direction 
of the external strain. Of these configurations, only the one with long axis along
the stretching direction of the external strain, could be attained 
without an internal control system.
In order for the extraction from the flow to take place, the microswimmer must
maintain on the average by an elongated shape in a direction between $-\pi/4$ 
and $\pi/4$ with respect to the flow.
The energy extracted from the flow that is not dissipated by internal friction,
could be converted into swimming strokes, (and thus returned to the external flow),
or in alternative, at least in principle, be stored in the swimmer,
in the form of some potential energy. 

All the configuration described above could be obtained by a system of brakes controlling
the stretching and contraction of the trimer arms in response to the external flow.
Inclusion of an elastic component in the dynamics may lead to higher swimming
efficiency: we have shown that, in the case of an ideal trimer dynamics, propulsion
requires a deformation component for energy extraction whose amplitude 
scales quadratically in the amplitude of the swimming stroke; internal 
dissipation would cause this scaling to become linear.

In the absence of an internal control system, a trimer, with dissipative 
springs between the beads, would be characterized,
in a viscous shear flow, by the same orientation pattern 
as a tank-treading vesicle \cite{keller82} or a microcapsule \cite{barthes80}.
The trimer would maintain, on the average, an elongated configuration, aligned
with the flow in the case of infinite friction, and aligned with the stretching
direction of the strain in the zero friction case.

The natural scale for the migration velocity of a microswimmer in an 
external flow is the external velocity difference $\alpha R$ across 
its body, where $\alpha$ is the shear strength [see Eq. (\ref{eq1})] and
$R$ is the swimmer size. The swimming velocity $U^{migr}$ 
of the swimmer in free space is going to be very small
$U^{migr}/(\alpha R)=O(a\delta R/R^2)$, where $a$ is the size of the moving
parts (the beads) and $\delta R$ is the amplitude of the swimming strokes. 
The correction by presence of a wall at distance $L$ from the swimmer is 
going to be smaller by an additional factor $(R/L)^2$.

Although small, the swimming velocity that can be achieved appear to dominate
Brownian diffusion as long as the swimmer size remains above the micron threshold.
Taking e.g. $R=10\mu{\rm m}$, $a=0.1R$ and $\delta R=0.3R$, with a shear strength
$\alpha=1{\rm s}^{-1}$, we find a migration velocity in free space 
$U^{migr}\sim 10^{-5}{\rm cm/s}$ [Eq. (\ref{eq17})] and a Brownian diffusivity
$D=KT/(6\pi\mu a)\sim 10^{-9}{\rm cm^2/s}$, corresponding to migration dominating
diffusion at distances above $R_c\sim 0.1R$ [Eq. (\ref{crossover})]. 
A swimmer that is ten times smaller,
would be characterized in the same flow by $U^{migr}\sim 10^{-6}{\rm cm/s}$
for a diffusivity $D\sim 10^{-8}{\rm cm^2/s}$ and a crossover length
$R_c\sim 10^2R$. Of course, the deterministic theory breaks down in this
regime, as $\Pe\sim 100$ and the contribution to trimer deformation from 
thermal noise becomes $O(1)$ [see Eqs. (\ref{peclet}) and (\ref{stroke})].
In this regime, however, rotational diffusivity of the trimer could be exploited
to allow propulsion with use of just one degree of freedom (but with an internal motor:
no more passive propulsion),
which provides another example of swimming strategy to which the 
limitations of the scallop theorem do not apply.

Coming back to the issue of swimming efficiency, as shown in \cite{olla10a}, 
it appears that a swimmer with a continuous body could achieve much better
results than the ones described in the present paper, namely, 
$O(\delta R/R)$ rather than $O(a\delta R/R^2)$ efficiency. Notice that this
is also better than the 
$O((\delta R/R)^2)$ result for a similar continuous swimmer in a quiescent 
fluid \cite{shapere89}. For $\delta R/R\sim 1$, the migration velocity
$U^{migr}$ $\sim\alpha R$ would have the necessary magnitude, to produce
phenomena, analogous to the Fahraeus-Lindqvist effect in small 
blood vessels \cite{vand48}.



\appendix
\setcounter{equation}{0}
\setcounter{section}{1}
\renewcommand{\theequation}{\thesection\arabic{equation}}
\section*{Appendix. The image field}
The image field must obey the equations of low Reynolds number hydrodynamics:
\beq
\nabla P=\mu\nabla^2\U^I,\quad
\nabla\cdot\U^I=0,
\label{Stokes}
\eeq
that is the Stokes equation plus incompressibility,
where 
$P$ is the pressure.
We can express $\U^I$
in terms of scalar and vector potentials:
\beq
\tilde\U^I=\nabla\Phi+\nabla\times\A,
\label{A1}
\eeq
where
\beq
\nabla^2\Phi=0
\quad
{\rm and}
\quad
\nabla\cdot\A=0.
\label{A2}
\eeq
The first of (\ref{A2}) is a consequence of incompressibility; the second 
is a gauge condition. From here, the vorticity equation $\nabla\times\nabla^2\tilde\U^I=0$,
which descends from Eq. (\ref{Stokes}),
can be written in the form
\beq
\nabla^2\nabla^2\A=0.
\label{A3}
\eeq
Fourier transforming with respect to $X_{2,3}$, 
the gauge condition can be written in the form
\beq
A_{2\k}=-\frac{k_3}{k_2}A_{3\k}+\frac{\im}{k_2}A'_{1\k},
\label{A4}
\eeq
where the prime indicates derivative with respect to $x_2$. The vorticity equation 
(\ref{A3}), instead, becomes $(\partial^2_{X_1}-k^2)^2\A_\k=0$, whose general solution
reads, imposing finiteness at $X_1\to -\infty$:
\beq
\A_\k(\X_1)&=&\hat\A_\k\,(X_1-L)\exp(k(X_1-L))
\nonumber 
\\
&+&{\bf a}_\k\exp(k(X_1-L)).
\label{A5}
\eeq
The second contribution to right hand side of Eq. (\ref{A5}) is a pure gauge term
that does not affect $\tilde\U^I$, and will be disregarded.
The first of Eqn. (\ref{A2}), instead, gives for $\Phi$:
\beq
\Phi_\k(\X_1)=\hat\Phi_\k\exp(k(X_1-L)).
\label{A6}
\eeq
Using Eqs. (\ref{A1}) and (\ref{A4}), the expression for the velocity correction becomes, 
in terms of Fourier components:
\beq
\begin{cases}
\tilde U^I_{1\k}={\im k^2\over k_2} A_{3\k}+{k_3\over k_2} A_{1\k}'+\Phi_\k',
\cr
\tilde U^I_{2\k}=\im k_3 A_{1\k}- A_{3\k}'+\im k_2\phi_\k,
\cr
\tilde U^I_{3\k}=-{k_3\over k_2} A_{3\k}'+{i\over k_2} A_{1\k}''-\im k_2
           A_{1\k}+ik_3\Phi_\k,
\end{cases}
\label{A7}
\eeq
and, imposing the boundary condition $\tilde\U^I_\k(L|0,\F_1)=-\tilde\U_\k(L|0,\F_1)$:
$$
\begin{cases}
\hat U_{1\k}=-{k_3\over k_2}\hat A_{1\k}-k\hat\Phi_\k,
\cr
\hat U_{2\k}=\hat A_3-\im k_2\hat\Phi_\k,
\cr
\hat U_{3\k}={k_3\over k_2}\hat A_{3\k}-{2\im k\over k_2}\hat A_{1\k}-\im k_3\hat\Phi_\k,
\end{cases}
$$
where $\hat\U_\k=\tilde\U_\k(L|0,\F_1)$ and use has been made of Eqs. (\ref{A5}-\ref{A6}).
Solution of this system gives:
\beq
\begin{cases}
\hat\Phi_\k=\frac{\im[-k_3k_2\hat U_{3\k}+2\im k_2k\hat U_{1\k}+k_3^2\hat U_{2\k}]}{2k_2k^2},
\cr
\hat A_{3\k}={k_3k_2\hat U_{3\k}-2\im k_2k\hat U_{1\k}+(k^2+k_2^2)\hat U_{2\k}\over 2k^2},
\cr
\hat A_{1\k}={\im (k_2\hat U_{3\k}-k_3\hat U_{2\k})\over 2k}.
\end{cases}
\label{A9}
\eeq
Substitution of Eqs. (\ref{A9}) together with Eqs. (\ref{A5}-\ref{A6}) into Eq. (\ref{A7}), gives,
at $X_1=0$:
\beq
&&\tilde U^I_{1\k}(0|0,\F_1)=-[(1+kL)\hat U_{1\k}
\nonumber
\\
&&+\im k_2L\hat U_{2\k}
+\im k_3L\hat U_{3\k}
]
\exp(-kL),
\nonumber
\eeq
where $\hat\U_\k\equiv\tilde\U_\k(L|0,\F_1)$.
In order to obtain $\Delta U^{migr}_1$, 
we thus have to antitransform
\beq
&&\partial_{X_1}\tilde U^I_{1\k}(0|\X,\F_1)|_{\X=0}=
[
(1+kL)\hat U'_{1\k}
\nonumber
\\
&&+\im k_2L\hat U'_{2\k}
+\im k_3L\hat U'_{3\k}
]
\exp(-kL),
\label{eq21}
\eeq
where $\hat\U'_\k\equiv \partial\hat\U_\k/\partial L$.
Since the trimer motion
is confined in the $X_1X_2$ plane, we do not need to calculate $\partial_{X_3}\tilde U^I_{1\k}$.
We thus get, antitransforming Eq. (\ref{eq21}) and
$\partial_{X_2}\tilde U^I_{1\k}(0|\X,\F_1)|_{\X=0}=-\im k_2\tilde U^I_{1\k}$ $(0|0,\F_1)$:
\beq
&&\partial_{X_1}\tilde U^I_1(0|\X,\F_1)|_{\X=0}
\nonumber
\\
&&=
\int\frac{\d^2k}{(2\pi)^2}\int\d^2Y_\perp\,
\exp(-\im\k\cdot\Y_\perp-kL)
\nonumber
\\
&&\times
[
(1+kL)\hat U'_1+\im k_2L\hat U'_2
+\im k_3L\hat U'_3
],
\label{A20}
\eeq
and
\beq
&&\partial_{X_2}\tilde U^I_1(0|\X,\F_1)|_{\X=0}
\nonumber
\\
&&=
\int\frac{\d^2k}{(2\pi)^2}\int\d^2Y_\perp\,
\exp(-\im\k\cdot\Y_\perp-kL)
\nonumber
\\
&&\times
[
\im(1+kL)\hat U_1-k_1L\hat U_2
- k_3L\hat U_3
]k_2,
\label{A21}
\eeq
where $\hat\U=\tilde\U(\Y|0,\F_1)$, 
$\Y=(L,\Y_\perp)$ and $\hat\U'=\partial\hat\U/\partial L$.
The stokeslet field at the wall $\hat\U$
is obtained from Eq.  (\ref{eq7}):
\beq
\hat U_1&=&\frac{3a}{4\sigma}\Big\{\frac{F_{11}}{(L^2+Y_\perp^2)^{1/2}}
+\frac{L[LF_{11}+Y_2 F_{12}]}{(L^2+Y_\perp^2)^{3/2}}\Big\},
\nonumber
\\
\hat U_2&=&\frac{3a}{4\sigma}\Big\{\frac{F_{12}}{(L^2+X_\perp^2)^{1/2}}
+\frac{Y_2[LF_{11}+Y_2 F_{12}]}{(L^2+Y_\perp^2)^{3/2}}\Big\},
\nonumber
\\
\hat U_3&=&\frac{3a}{4\sigma}
\frac{Y_3[LF_{11}+Y_2 F_{12}]}{(L^2+Y_\perp^2)^{3/2}}.
\nonumber
\label{A22}
\eeq
The integrals in Eqs. (\ref{A20}-\ref{A21}) are carried out in polar coordinates
with the help of MAPLE, and the result 
is Eq. (\ref{eq22}).


\begin{thebibliography}{10}
\bibitem{note} In the case of a discrete swimmer, it would be perhaps more
appropriate to speak of pseudo-tank-treading motion, as discrete tank-treading
goes as tank-treading, in the same way as, say in a car, a triangular wheel would go 
as a round wheel.
\bibitem{note1} The contribution to diffusivity from random swimming can 
be estimated as $D_u\sim\alpha^{-1}\langle(\delta U^{migr})^2\rangle$,
where $\delta U^{migr}\sim\alpha a\delta r/R$. [We are maximizing $D_u$,
taking $\alpha^{-1}$ as characteristic time for $\delta U^{migr}$; 
see Eq. (\ref{stroke})]. 
Notice that rotational diffusivity
is not a problem as long as the trimer is able to modify its shape in 
response to orientation. 
We see that $D/D_u\sim\alpha\tau_c$ and random swimming will dominate
for $\alpha\tau_c\ll 1$. In this case, however, $R_c\sim D_u/U^{migr}$
and from Eq. (\ref{alphatau_c}): $R_c/R\sim za/R\ll 1$.
\end{thebibliography}
\end{document}